\documentclass[review]{biometrika}
\usepackage{amsmath, amsfonts, graphicx, amssymb, setspace, caption, subcaption, color}
\usepackage{times}
\usepackage{bm}
\usepackage{natbib}

\usepackage[plain,noend]{algorithm2e}

\makeatletter
\renewcommand{\algocf@captiontext}[2]{#1\algocf@typo. \AlCapFnt{}#2} % text of caption
\def\@algocf@capt@plain{top}
\renewcommand{\algocf@makecaption}[2]{%
  \addtolength{\hsize}{\algomargin}%
  \sbox\@tempboxa{\algocf@captiontext{#1}{#2}}%
  \ifdim\wd\@tempboxa >\hsize%     % if caption is longer than a line
    \hskip .5\algomargin%
    \parbox[t]{\hsize}{\algocf@captiontext{#1}{#2}}% then caption is not centered
  \else%
    \global\@minipagefalse%
    \hbox to\hsize{\box\@tempboxa}% else caption is centered
  \fi%
  \addtolength{\hsize}{-\algomargin}%
}
\makeatother

\newcommand{\rr}{{r}}

\newcommand{\X}{X}

\renewcommand{\x}{\vec{x}}
\newcommand{\E}{\mathbb{E}}
\newcommand{\bTheta}{{ \Theta }}

\newcommand{\Minimize}{\operatornamewithlimits{minimize}}
\newcommand{\argmin}{\operatornamewithlimits{arg\,min}}

\newcommand{\bPsi}{ {\Psi}}
\newcommand{\bSigma}{ { \Sigma}}

\begin{document}
%\jname{Biometrika}
%% The year, volume, and number are determined on publication
\jyear{2013}
%\jvol{99}
%\jnum{1}
%% The \doi{...} and \accessdate commands are used by the production team
%\doi{10.1093/biomet/asm023}
%\accessdate{Advance Access publication on 31 July 2012}
%\copyrightinfo{\Copyright\ 2012 Biometrika Trust\goodbreak {\em Printed in Great Britain}}

%% These dates are usually set by the production team
\received{April 2013}
%\revised{September 2012}

%% The left and right page headers are defined here:
\markboth{A. Voorman, A. Shojaie, D. Witten}{Biometrika}

%% Here are the title, author names and addresses
\title{Graph Estimation with Joint Additive Models}

\author{Arend Voorman, Ali Shojaie, and Daniela Witten}
\affil{Department of Biostatistics, University of Washington  \email{voorma@uw.edu} }

\maketitle

\begin{abstract}
In recent years, there has been considerable interest in estimating conditional independence graphs in the high-dimensional setting.  Most prior work has assumed that the variables are multivariate Gaussian, or that the conditional means of the variables are linear. Unfortunately, if these assumptions are violated, then the resulting conditional independence estimates can be inaccurate. We present a semi-parametric method, SpaCE JAM, which allows the conditional means of the features to take on an arbitrary additive form. We present an efficient algorithm for its computation, and prove that our estimator is consistent.  We also extend our method to estimation of  directed graphs with known causal ordering. Using simulated data, we show that SpaCE JAM enjoys superior performance to existing methods when there are non-linear relationships among the features, and is comparable to methods that assume multivariate normality when the conditional means are linear. We illustrate our method on a cell-signaling data set.
\end{abstract}

\begin{keywords}
graphical models; sparse additive models; lasso; sparsity; conditional independence; nonlinearity; non-Gaussianity\vspace{-40pt}

\end{keywords}
\setstretch{2}
\section{Introduction}\label{sec:intro}
In recent years, there has been considerable interest in developing methods to estimate the joint pattern of association among a set of random variables.  The relationships between $d$ random variables can be summarized with an undirected graph $\Gamma = (V,E)$ in which the random variables are represented by the vertices $V = \{1,\ldots,d\}$ and the conditional dependencies between pairs of variables are represented by edges $E \subset V \times V$. That is, for each $j \in V$, we want to determine a minimal set of variables on which the conditional densities $p_j(x_j \mid \{x_k, k \neq j\})$ depend,
$$p_j(x_j \mid \{x_k, k \neq j\}) = p_j(x_j \mid \{x_k: (k,j) \in E \}).$$
{ Recently there has also been considerable work in estimating  marginal associations between a set of random variables \citep[see e.g.][]{basso2005reverse,meyer2008minet,kuo2008gene,hausser2009entropy,chen2010nonparametric};
however, in this paper we focus on  conditional dependencies, which provide richer information about the relationships among the variables. }% However, in this paper we focus on  the conditional distributions, as they model the joint effect of all features on each other.}

\begin{figure}
        \centering
	\includegraphics[width =\textwidth]{\detokenize{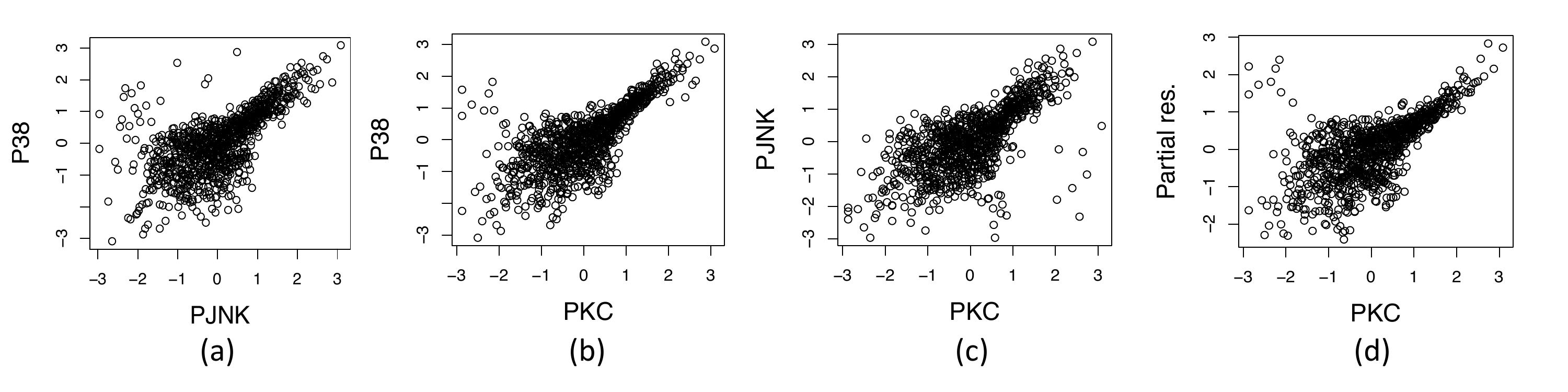}}
        \caption{Cell signaling data from \citet{sachs2005causal}. (a)-(c) Pairwise scatterplots for PKC, P38 and PJNK. (d) Partial residuals from the linear regression of P38 on PKC and PJNK.  The data are standardized to have normal marginal distributions, but are clearly not multivariate normal.}
\label{fig:example1}
\vspace{-20pt}
\end{figure}

Estimating the conditional independence graph $\Gamma$ based on a set of $n$ observations is an old problem \citep{dempster1972}. % that has been the subject of renewed interest in recent years.
In the case of high-dimensional continuous data, most  prior work  has assumed either (a) multivariate Gaussianity~\cite[see e.g.][]{friedman2008,rothman2008sparse,yuan2007,banerjee2008} or (b) linear conditional means~\cite[see e.g.][]{meinshausen2006,peng2009} for the features.
However, as we will see, these two  assumptions are essentially equivalent.
As an illustration, consider the cell signaling data set from \citet{sachs2005causal}, which consists of protein concentrations measured under a set of perturbations. We analyze the data set in more detail in Section~\ref{sec:realdata}. Pairwise scatterplots of three of the variables are given in Figure \ref{fig:example1} (a)-(c) for one of 14 perturbations.  Here, the data have been transformed to be marginally normal, as suggested by \citet{liu2009nonparanormal}. The transformed data clearly are not multivariate normal, given the non-constant variance in the bivariate scatterplots, and as confirmed by a Shapiro-Wilk test ($p<2 \times 10^{-16}$). % \cite[$p < 2\times 10^{-16}$,][]{shapiro1965analysis}.

Can the data in Figure~\ref{fig:example1} be well-represented by linear relationships? In Figure~\ref{fig:example1} (d), we see strong evidence that the conditional mean of the protein P38 given PKC and PJNK is nonlinear.  This is corroborated by the fact that the $p$-value for including quadratic terms in the linear regression of P38 onto PKC and PJNK is small $(p < 2 \times 10^{-16})$. Therefore in this data set, the features are not multivariate Gaussian, and marginal transformations do not remedy the problem.

In order to flexibly model conditional mean relationships, we could specify a more flexible joint distribution.  However, joint distributions are difficult to construct and computationally challenging to fit, and the resulting conditional models need not be easy to obtain or interpret. Alternatively we can specify the conditional distributions directly. This has the advantage of simpler interpretation and greater computational tractability. In this paper, we will model the conditional means of non-Gaussian random variables  with generalized additive models \citep{hastie1990generalized}, and will use these in order to construct conditional independence graphs.

Throughout this paper, we will assume that we are given $n$ independent and identically distributed observations from a $d$-dimensional random vector $x=(x_1, \dots, x_d) \sim \mathcal{P}$. Our observed data can be written as $X = [ \x_1, \ldots  \x_d] \in \mathbb{R}^{n \times d}$.

The rest of the paper is organized as follows.  In Sections~\ref{sec:background} and \ref{sec:previous} we review methods for modeling conditional dependence relationships among a set of variables, and discuss their limitations. In Section~\ref{sec:method} we propose our method (SpaCE JAM) and an algorithm for its computation. We illustrate our method on real and simulated data in Section~\ref{sec:experiments}, and compare with available methods.  In Section~\ref{sec:dags} we extend the method to the estimation of directed acyclic graphs with known causal ordering.  In Section~\ref{sec:theory} we prove consistency of our algorithm, and in Section~\ref{sec:extension} we propose a screening rule for estimation in high dimensions. The discussion is in Section~\ref{sec:discussion}.
\vspace{-30pt}
\section{Modeling conditional dependence relationships}
\label{sec:background}
%Suppose we are interested in estimating the conditional independence graph for a random vector $X \sim \mathcal{P}_\theta$ for some family of distributions indexed by $\theta$: $p(X; \theta) = p(x_1, \dots, x_d; \theta)$.
%The joint distribution completely specifies the conditional distributions. That is, if $p_j$ is the conditional distribution of $x_j \mid \{x_k : k \neq j\}$, then
%$$\log(p(X; \theta))  \propto \phi(X; \theta) + \sum_{j=1}^d \log( p_j(x_j \mid \{x_k: (j,k) \in E \}; \theta)),$$
%where $\phi(\cdot)$ is a normalizing term. In order to estimate $\Gamma$ it suffices to obtain an estimate $\hat \theta$ for $\theta$, using e.g. maximum likelihood.

{Suppose we are interested in estimating the conditional independence graph $\Gamma$ for a random vector $x \in \mathbb{R}^d$. If the joint distribution is known up to some finite dimensional parameter $\theta$, then to estimate $\Gamma$ it suffices to estimate $\theta$ via e.g. maximum likelihood.}
One practical difficulty that arises in estimating $\Gamma$ is specification of a plausible joint distribution. Specifying a conditional distribution, such as in a regression model, is typically much less daunting. We therefore consider pseudo-likelihoods \citep{besag1974spatial, besag1975statistical} of the form
$$\log(p_{PL}(x; \theta)) = \sum_{j=1}^d  \log\left( p_j(x_j \mid \{x_k : (j,k) \in E \}; \theta)\right).$$
For a set of arbitrary conditional distributions, there need not be a compatible joint distribution \citep{wang2008conditionally}. However, the conditionally specified graphical model has an appealing theoretical justification, in that it minimizes the Kullback-Leibler distances to the conditional distributions \citep{varin2005note}. Furthermore, in estimating conditional independence graphs, our scientific interest is in the conditional independence relationships rather than in the joint distribution. So in a sense, modeling the conditional distribution rather than the joint distribution amounts to a more direct approach to graph estimation. We therefore advocate for an approach for non-Gaussian graphical modeling based on conditionally specified models \citep{varin2011overview}.
\vspace{-40pt}
\section{Previous work}%%%%%%%%%%%%%%%%%%%%%%%%%%%%%%%%%%%%%%%%%%%%%%%%%%%%%%%%%%%%%%%%%
\label{sec:previous}
\vspace{-10pt}
\subsection{Estimating graphs with Gaussian data}
\label{sec:gauss}
Suppose for now that $x$ has a joint Gaussian distribution with mean 0 and precision matrix $\bTheta$. One can write the {negative} log-likelihood of the joint distribution, up to constants, as
\begin{equation}\label{mvn}
-\log \det({ \Theta})  + \mathrm{tr}\left(xx^T {\Theta} \right).
\end{equation}
In this case, the conditional relationships are linear,
\begin{equation}
\label{linmod}
x_j \mid \{x_k, \, k \neq j\} = \sum_{k \neq j} \beta_{jk} x_k +\epsilon_j, \quad j =1,\dots,d,
\end{equation}
where $\beta_{jk} = -\bTheta_{jk}/\bTheta_{kk}$ and $\epsilon_j \sim N_1(0,1/\bTheta_{jj})$. To estimate the graph $\Gamma$, we must determine which $\beta_{jk}$ are zero in (\ref{linmod}), or equivalently which $\bTheta_{jk}$ are 0 in (\ref{mvn}). This is simple when $n \gg d$.

In the high-dimensional setting, when the maximum likelihood estimate is unstable or undefined, a number of approaches have been proposed to estimate the conditional independence graph $\Gamma$, which we review here.
%Using the conditional framework,
 \citet{meinshausen2006} proposed fitting (\ref{linmod}) using an $\ell_1$-penalized regression. This is  referred to as neighborhood selection:
\begin{equation}
\left \{\hat \beta_{jk}: 1 \leq j,k \leq d\right \} = \argmin_{\beta_{jk}: 1 \leq j,k \leq d} \left\{   \frac{1}{2}\sum_{j=1}^d  \|\x_j - \sum_{k \neq j} \x_k \beta_{jk}\|^2 + \lambda \sum_{j=1}^d \sum_{k \neq j} | \beta_{jk}|     \right\}.
\label{MB}
\end{equation}
Here $\lambda$ is a nonnegative tuning parameter that encourages sparsity in the  coefficient estimates.  \citet{peng2009} improved upon the neighborhood selection approach by applying $\ell_1$ penalties to the partial correlations; this
%$\rho_{jk} = \beta_{jk} \left(\bTheta_{kk}/\bTheta_{jj}\right)^{1/2} = \beta_{kj} \left(\bTheta_{jj}/\bTheta_{kk}\right)^{1/2}$. This
 is known as sparse partial correlation estimation.

As an alternative to (\ref{MB}), many authors have considered estimating $\bTheta$ under the multivariate normality assumption by maximizing an $\ell_1$-penalized joint log likelihood \citep[see e.g.][]{yuan2007,banerjee2008,friedman2008}.
This amounts to the optimization problem
\begin{equation}
 \label{glasso}
\hat \bTheta = \argmin_{{\Theta} \succ 0}  \left\{ -\log \det( \Theta)  + \mathrm{tr}\left(\X^T\X  \Theta \right)/n + \lambda \| \Theta\|_1 \right \},
\end{equation}
known as the graphical lasso.
The solution $\hat{\Theta}$ to (\ref{glasso}) serves as an estimate for $\bTheta$, and hence the sparsity pattern of $\hat \Theta$ (induced by the $\ell_1$ penalty) provides an estimate of $\Gamma$.

At first glance, neighborhood selection and sparse partial correlation may seem semi-parametric: a linear model may hold in the absence of multivariate normality. However, while (\ref{linmod}) can accurately model each conditional dependence relationship semi-parametrically, the accumulation of these specifications is very restrictive in terms of the joint distribution.  In fact, \citet{khatri1976characterizations} proved that if (\ref{linmod}) holds, along with some other mild assumptions, then the joint distribution must be multivariate normal. Notably, this is true regardless of the distribution of the errors $\epsilon_1,\ldots,\epsilon_d$ in (\ref{linmod}). In other words, even though  (\ref{MB}) does not explicitly involve the multivariate normal likelihood, normality is implicitly assumed. This means that if we wish to model non-normal continuous data, then non-linear conditional models are necessary.

%It has been shown that \eqref{MB} is an approximation to \eqref{glasso} \citep{friedman2008}. For instance, when $\lambda=0$, it is not hard to show that for a given $j$, the $(j,k)$ element of $\hat{\bf \Theta}$ in \eqref{glasso} is proportional to $\hat\beta_{jk}$ in \eqref{MB} for all $k \neq j$.
%%%%%%%%%%%%%%%%%%%%%%%%%%%%%%%%%%%%%%%%%%%%%%%%%%%%%%%%%%%%%%%%%
\vspace{-25pt}
\subsection{Estimating graphs with non-Gaussian data}
\label{sec:nongauss}
We now briefly review three existing methods for modeling conditional independence graphs with non-Gaussian data. The normal copula or nonparanormal model
(\citealt{liu2009nonparanormal}, \citealt{liu2012high}, \citealt{xue2012regularized}, { studied in the Bayesian context by \citealt{dobra2011copula}}) assumes that  $x$ has a nonparanormal distribution: that is,  $(h_1(x_1), \dots, h_d(x_d)) \sim N_d(0, \bTheta)$ for functions $h_1(\cdot),\ldots,h_d(\cdot)$.  After $h_1(\cdot),\ldots,h_d(\cdot)$ are estimated, one can apply any of the methods mentioned  in Section~\ref{sec:gauss} to the transformed data.
The conditional model implicit in this approach is
\begin{equation}
\label{npnmod}
h_j(x_j) \mid \{x_k, \, k \neq j\} = \sum_{k \neq j} \beta_{jk} h_k(x_k) + \epsilon_j, \quad j =1,\dots,d.
\end{equation}
This is itself a restrictive assumption, which may not hold, as seen in Figure~\ref{fig:example1}.

Forest density estimation \citep{liu2011forest} replaces the need for distributional assumptions with graphical assumptions: the underlying graph is assumed to be a forest. Then bivariate densities are estimated non-parametrically. Unfortunately, the restriction to acyclic graphs may be inappropriate in applications, and maximizing over all possible forests is infeasible.

{The graphical random forests \citep{fellinghauer2011stable} approach uses random forests to flexibly model conditional means, and allows for interaction terms. But this  does not correspond to a well-defined statistical model, and guarantees on feature selection consistency are unavailable.}
%%%%%%%%%%%%%%%%%%%%%%%%%%%%%%%%%%%%%%%%%%%%%%%%%%%%%%%%%%%%%%%%%
%%%%%%%%%%%%%%%%%%%%%%%%%%%%%%%%%%%%%%%%%%%%%%%%%%%%%%%%%%%%%%%%%
\vspace{-40pt}
\section{Method}
\label{sec:method}
\subsection{Jointly additive models}
In order to estimate a conditional independence graph using a pseudolikelihood approach, we must estimate the variables on which the conditional distributions $p_j(\cdot)$ depend. However, since density estimation is generally a challenging task, especially in high dimensions, we focus on the simpler problem of estimating the conditional mean $\E[\,x_j \mid \{ x_k: (j,k)  \in E\}\,]$, under the assumption that the conditional distribution and the conditional mean depend on the same set of variables.  Thus, we seek to estimate the conditional mean $f_j(\cdot)$ in the regression model
$$x_j | \{x_k, k \neq j\} = f_j\left(x_k:  k \neq j \right) +\epsilon_j,$$
where $\epsilon_j$ is a mean-zero error term. Since estimating arbitrary functions $f_j(\cdot)$ is infeasible in high dimensions, we restrict ourselves to additive models of the form
\begin{equation}
x_j  | \{x_k, \, k \neq j\} = \sum_{k \neq j} f_{jk} (x_k) + \epsilon_{j},
\label{gamus}
\end{equation}
where $f_{jk}(\cdot) \in \mathcal{F}$ for some space of functions $\mathcal{F}$. This amounts to modeling each variable  using a generalized additive model \citep{hastie1990generalized}.   {Unlike  \citet{fellinghauer2011stable}, we do not assume that the errors $\epsilon_{j}$ are independent of the additive components $f_{jk} (\cdot)$, but merely that the conditional independence structure can be recovered from the additive components $f_{jk} (\cdot)$.}

%{We need not beleive that the model is truly additive, here we simply state that we are interested in  additive approximations. Thus, we allow for the error terms $\epsilon_{j}$ to contain both independent noise and approximation error.}
%%%%%%%%%%%%%%%%%%%%%%%%%%%%%%%%%%%%%%%%%%%%%%%%%%%%%%%%%%%%%%%%%
%\subsection{Estimation in the low-dimensional setting}
%%%%%%%%%%%%%%%%%%%%%%%%%%%%%%%%%%%%%%%%%%%%%%%%%%%%%%%%%%%%%%%%%
\subsection{Estimation with SpaCE JAM}
Since we believe that the conditional independence graph is sparse, we fit \eqref{gamus} using a penalty that performs simultaneous estimation and selection of the $f_{jk}(\cdot)$. Specifically, we link together $d$ sparse additive models
\citep{ravikumar2009sparse}
using a penalty
%use a variation %estimate each variable in turn
%This can be done for each variable in turn
%using a variation of
%of a  sparse additive model \citep{ravikumar2009sparse}
 that groups  the parameters corresponding to a single edge in the graph.
This results in the problem
%gives rise
%. Here we propose to group the parameters in a sparse additive model according to edges in a graph, giving rise
% to the optimization problem
\begin{equation}
\Minimize_{f_{jk} \in \mathcal{F}, 1 \leq j,k \leq d}\left \{ \frac{1}{2n} \sum_{j=1}^d \| \x_j - \sum_{k \neq j} f_{jk}(\x_k)\|_2^2 + \lambda \sum_{k >j} \left(\| f_{jk}(\x_k) \|_2^2 + \|f_{kj}(\x_j)\|_2^2  \right)^{1/2} \right \}.
\label{eqn:us}
\end{equation}
We consider $f_{jk}(\x_k) = \bPsi_{jk} \beta_{jk}$, where $\bPsi_{jk}$ is a $n \times r$ matrix whose columns are basis functions used to model the additive components $f_{jk}$, and  $\beta_{jk}$ is an $r$-vector containing  the associated coefficients.  For instance, if we  use a linear basis function, i.e. $\bPsi_{jk} = \x_k$, then $r = 1$ and we are modeling only linear conditional means, as in \citet{meinshausen2006}. Higher-order terms allow us to model more complex dependencies.  The standardized group lasso penalty \citep{simon2011standardization} encourages sparsity and ensures that  the estimates of $f_{jk}(\cdot)$ and $f_{kj}(\cdot)$ will be simultaneously zero or non-zero.
Problem (\ref{eqn:us}) is the natural extension of  sparse additive modeling \citep{ravikumar2009sparse} to graphs, and generalizes neighborhood selection \citep{meinshausen2006} and sparse partial correlation \citep{peng2009} to allow for  flexible conditional means.  We call the solution to (\ref{eqn:us}) SpaCE JAM  (for SPArse Conditional Estimation with Joint Additive Models), to reflect its ties with the aforementioned techniques.
\begin{algorithm}[!h]
\caption{SpaCE JAM algorithm}\label{alg:spacejam}
\begin{tabbing}
Initialize $\hat \beta$'s \\
Repeat until convergence: \\
\enspace For $(j,k) \in V \times V$: \\
\quad 1: Calculate the vector of residuals for the $j$th and $k$th variables: \\
\quad \quad  $ \rr_{jk} \gets \x_j - \sum_{i \neq j,k} \bPsi_{ji} \hat{ \beta}_{ji}$\\
\quad \quad  $ \rr_{kj} \gets \x_k - \sum_{i \neq j,k} \bPsi_{ki} \hat{ \beta}_{ki}$ \\	
\quad  2: Regress the residuals on the specified basis functions:\\
\quad \quad $ \hat{ \beta}_{jk} \gets  \left(\bPsi_{jk}^T\bPsi_{jk}\right)^{-1}\bPsi_{jk}^T\rr_{jk}$ \\
\quad \quad $ \hat{ \beta}_{kj} \gets \left(\bPsi_{kj}^T\bPsi_{kj}\right)^{-1}\bPsi_{kj}^T\rr_{kj} $\\
\quad 3: Threshold:\\
\quad \quad  $\hat{ \beta}_{jk} \gets \left(1-n\lambda\left( \| \bPsi_{jk} \hat{\beta}_{jk}\|_2^2 + \| \bPsi_{kj} \hat{ \beta}_{kj}\|_2^2\right)^{-1/2}  \right)_+\hat{ \beta}_{jk} $ \\
\quad \quad  $\hat{ \beta}_{kj} \gets \left(1-n\lambda\left(\| \bPsi_{jk} \hat{ \beta}_{jk}\|_2^2 + \| \bPsi_{kj} \hat{\beta}_{kj}\|_2^2\right)^{-1/2}  \right)_+\hat{ \beta}_{kj} $\\
\end{tabbing}
\vspace{-30pt}
\end{algorithm} \newline Algorithm~\ref{alg:spacejam} uses block coordinate descent to solve (\ref{eqn:us}). Since \eqref{eqn:us} is convex,  the algorithm converges to the global minimum \citep{simon2011standardization}.
Performing Step 2  requires an $r \times r$ matrix inversion, where $r$ is the number of basis functions; this must be performed only  twice per pair of variables.
Estimating 30 conditional independence graphs with $r=3$ on a simulated data set with $n=50$ and $d=100$ takes  1.1 seconds on a 2.8 GHz Intel Core i7 Macbook Pro.
\vspace{-10pt}
{\subsection{Tuning}
A number of options for tuning parameter selection are available,
such as generalized cross-validation \citep{tibs1996}, the Bayesian
information criterion \citep{zou2007degrees}, and stability selection \citep{meinshausen2010stability}. We take an approach motivated by the Bayesian information criterion, as in  \citet{peng2009}. For the $j$th variable, the criterion is
\begin{equation}
\label{bic}
\textsc{bic}_j(\lambda) = n\log(\textsc{rss}_j (\lambda)) + \log(n)\textsc{df}_j(\lambda),
\end{equation}
where $\textsc{rss}_j (\lambda) = \| \x_j - \sum_{k \neq j} \bPsi_{jk} \hat{ \beta}^{(\lambda)}_{jk}\|_2^2$ is the residual sum of squares from minimizing \eqref{eqn:us} with tuning parameter $\lambda$, and $\textsc{df}(\lambda)_j$ is the degrees of freedom used in this regression.
We seek the value of $\lambda$ that minimizes
$\sum_{j=1}^d \textsc{bic}_j(\lambda)$.
When a single basis function is used, we can approximate the degrees of freedom by the number
of non-zero parameters in the regression
\citep{zou2007degrees,peng2009}. But when $r>1$  basis functions are used,  we use
\begin{equation}
\textsc{df}_j(\lambda)= |S^{(\lambda)}_j| + (r-1) \sum_{k}\frac{\|\bPsi_{jk} \hat{ \beta}^{(\lambda)}_{jk}\|_2^2}{\|\bPsi_{jk} \hat{ \beta}^{(\lambda)}_{jk}\|_2^2+ \lambda}, \label{df}
\end{equation}
where $S^{(\lambda)}_j=\{k : \|\hat \beta^{(\lambda)}_{jk}\| \neq 0\}$. Though (\ref{df}) was derived under the assumption of an orthogonal
design matrix,  it is a good approximation for the
non-orthogonal case \citep{yuan2006model}.}

{In order to perform SpaCE JAM, we must select a set of basis functions. In the absence of domain knowledge, we use cubic polynomials, which can approximate a wide range of functions.}
%}
%%%%%%%%%%%%%%%%%%%%%%%%%%%%%%%%%%%%%%%%%%%%%%%%%%%%%%%%%%%%%%%%%
\vspace{-40pt}
\section{Numerical experiments}
\label{sec:experiments}
%%%%%%%%%%%%%%%%%%%%%%%%%%%%%%%%%%%%%%%%%%%%%%%%%%%%%%%%%%%%%%%%%
\subsection{Simulation setup}
\label{sec:sim}
As discussed in Section~\ref{sec:background}, it can be  difficult to specify flexible non-Gaussian distributions for continuous variables.  However, construction of  multivariate distributions via conditional distributions is straightforward when the variables can be represented with a directed acyclic graph. The joint probability distribution of variables in a directed acyclic graph can be decomposed as $p(x_1, \dots, x_d) = \prod_{j =1}^d p_j(x_j | \{x_k :  (k,j) \in E_{D}\}),$
where $E_D$ denotes the {directed} edge set of the graph. {This is a valid joint distribution regardless of the choice of conditional distributions} $p_j(x_j | \{x_k :  (k,j) \in E_{D}\})$ \citep[][Chapter $1.4$]{pearl2000causalityStruct}.
We chose structural equations of the form
\begin{equation}\label{eqn:gamdag}
x_j | \{x_k :  (k,j) \in E_{D}\} =  \sum_{(k,j) \in E_D } f_{jk}(x_k)+ \epsilon_j,
\end{equation}
with $\epsilon_j \sim N(0,1)$.  If  the $f_{jk}$ are chosen to be linear, then the data are multivariate normal, and if the $f_{jk}$ are non-linear, then the data will typically not correspond to a well-known multivariate distribution.
We moralized the directed graph in order to obtain the
conditional independence graph  \citep[][Chapter $3.2$]{cowell2007probabilisticMoral}. {Note that here we have used directed acyclic graphs simply as a tool to generate non-Gaussian data, and that the full conditional distributions of the random variables created using this approach are not necessarily additive.}

We first generated a directed acyclic graph with $d=100$ nodes and 80 edges chosen at random from the ${100 \choose 2}$ possible edges. We used two schemes to construct a distribution on this graph. In the first setting, we chose $f_{jk}(x_k) = b_{jk1}x_k + b_{jk2} x_k^2 + b_{jk3}x_k^3, $
%where $\x_k^q$ is a $n$-vector whose $i$th element is $X^q_{ik}$, and
where the $b_{jk1}$, $b_{jk2}$, and $b_{jk3}$ are independent and normally distributed with mean zero and variance $1$, $0.5$, and $0.5$, respectively. In the second case, we chose $f_{jk}(\x_k)=\x_k$, resulting in multivariate normal data. In both cases we scaled the  $f_{jk}(\x_k)$ to have unit variance.
%\vspace{-20pt}
We generated  $n=50$ observations, and compared  SpaCE JAM to sparse partial correlation \citep[][\texttt{R} package \texttt{space}]{peng2009},  graphical lasso \citep[][\texttt{R} package \texttt{glasso}]{yuan2007}, neighborhood selection \citep[][\texttt{R} package \texttt{glasso}]{meinshausen2006},  nonparanormal \citep[][\texttt{R} package \texttt{glasso}]{liu2012high,xue2012regularized},   forest density estimation \citep[][code provided by authors]{liu2011forest}, the method of \citet[][\texttt{R} package \texttt{minet}]{basso2005reverse}, and graphical random forests \citep[][code provided by authors]{fellinghauer2011stable}.
% We used the \verb!R! packages \verb!space! to implement sparse partial correlation, \verb!glasso! for the graphical lasso, neighborhood selection and the nonparanormal skeptic, \verb!minet! for \citet{basso2005reverse}, and code from the authors for forest density estimation and graphical random forests.
%using the \verb!R! packages \verb!glasso!, \verb!space! and \verb!huge! to implement the graphical lasso, sparse partial correlation, and neighborhood selection and the nonparanormal transformation, respectively.
In performing neighborhood selection, we declared an edge between the $j$th and $k$th variables if
$\hat\beta_{jk} \neq 0$ or $\hat\beta_{kj} \neq 0$.   We performed  SpaCE JAM using three sets of basis functions: $\bPsi_{jk} = [\, \x_k, \x_k^2\,]$, $\bPsi_{jk} = [\, \x_k, \x_k^3\,]$, and $\bPsi_{jk} = [\, \x_k, \x_k^2, \x_k^3\,]$.

\subsection{{Simulation results}}

\begin{figure}
\includegraphics[width=\textwidth]{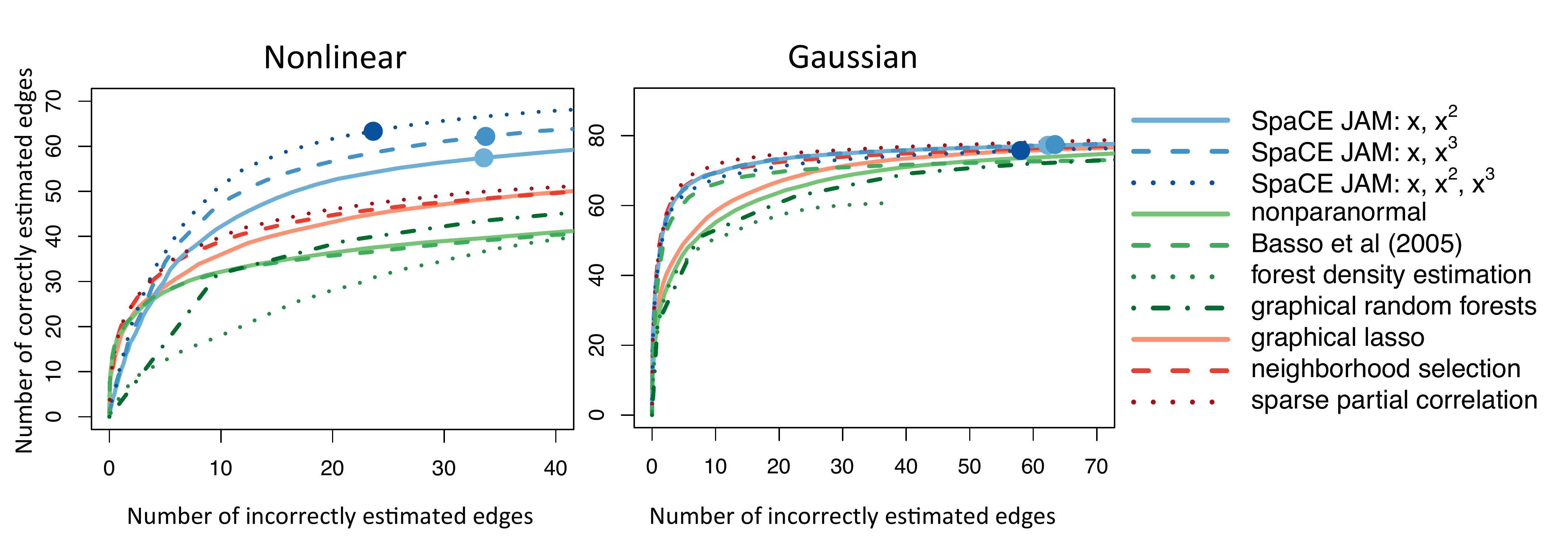}
\caption{Simulation study. The number of correctly estimated edges is displayed as a function of incorrectly estimated edges, for a range of tuning parameter values, in the non-linear (left) and Gaussian (right) set-ups, averaged over 100 simulated data sets.  Dots indicate the average model size chosen using the $\textsc{bic}$ criterion. In the order of appearance in the legend, the competing methods are those of \citet{liu2012high,basso2005reverse,liu2011forest,fellinghauer2011stable,yuan2007,meinshausen2006,peng2009}.}
\label{simfig}
\vspace{-10pt}
\end{figure}

Figure~\ref{simfig} summarizes the results of our simulations. For each method, the numbers of correctly and incorrectly estimated edges were averaged  over 100 simulated data sets for a range of 100 tuning parameter values. When the $f_{jk}(\cdot)$ are non-linear, SpaCE JAM with the basis $\bPsi_{jk} = [\, \x_k, \x_k^2,\x_k^3\,]$ dominates SpaCE JAM with the basis sets $\bPsi_{jk} = [\, \x_k, \x_k^2\,]$ or $ [\, \x_k, \x_k^3\,]$, which in turn tend to enjoy superior performance relative to all other methods (left panel of Figure~\ref{simfig}). Furthermore, even though the basis sets $\bPsi_{jk} = [\, \x_k, \x_k^2\,]$ and $ [\, \x_k, \x_k^3\,]$ do not entirely capture the functional forms of the data-generating mechanism, they still outperform methods that assume linearity, as well as competitors intended to model non-linear relationships. % The nonparanormal has a higher false positive rate than the graphical lasso, despite the former's goal of effectively modeling non-linear  conditional means. %\\

When the conditional means are linear and the number of estimated edges is small, all methods perform roughly equally (right panel of Figure~\ref{simfig}).   {As the number of estimated edges is increased}, sparse partial correlation performs best, 
%while the graphical lasso and the nonparanormal perform worse. 
while the graphical lasso, the nonparanormal and the forest-based methods perform worse.
This agrees with the observations of  \citet{peng2009} that sparse partial correlation and neighborhood selection tend to outperform the graphical lasso.  In this setting, since non-linear terms are not needed to model the conditional dependence relationships, sparse partial correlation outperforms SpaCE JAM with two basis functions, which performs better than SpaCE JAM with three basis functions. Nonetheless, the loss in accuracy due to the inclusion of non-linear basis functions is not dramatic, and SpaCE JAM still tends to outperform other methods for non-Gaussian data, as well as the graphical lasso.

\vspace{-20pt}
\subsection{Application to cell signaling data}\label{sec:realdata}
\begin{figure}[h!]
\begin{center}
\vspace{-5pt}
\includegraphics[width=0.8\textwidth]{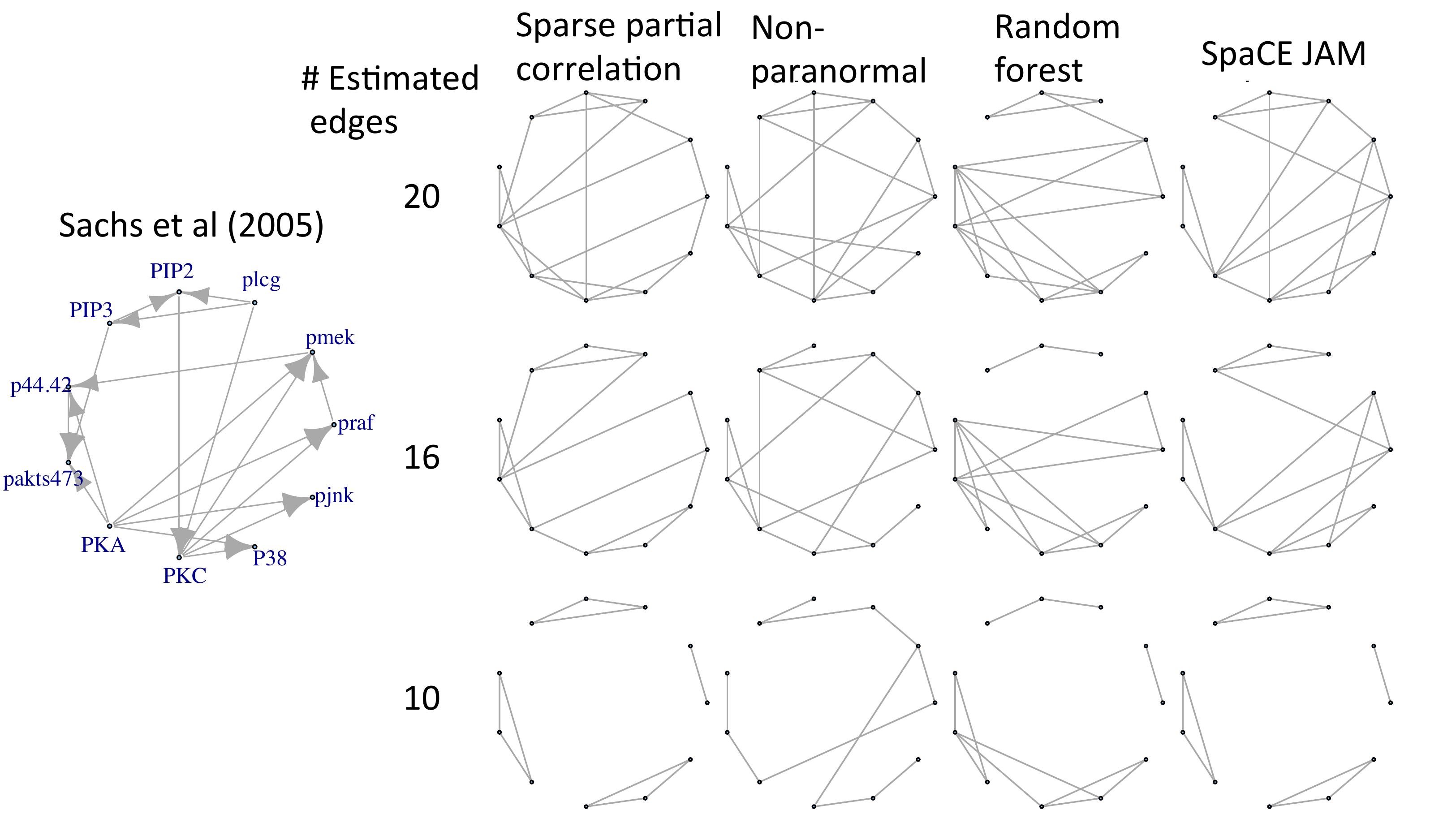}
\caption{Cell signaling  data set; graph reported in \citet{sachs2005causal} is shown on the left.  On the right, graphs were estimated using data from one perturbation of the data set.  From top to bottom, panels contain graphs with 20, 16 and 10 edges.  From left to right, comparisons are to \citet{peng2009,liu2012high,fellinghauer2011stable}. We cannot specify an arbitrary graph size using graphical random forests,  so graph sizes for that approach do not match exactly.}
\label{cellsfig}
\end{center}
\vspace{-20pt}
\end{figure}

We apply  SpaCE JAM to a data set consisting of measurements for 11 proteins involved in cell signaling, under 14 different perturbations \citep{sachs2005causal}. To begin, we consider data from one of the 14 perturbations ($n=911$), and compare SpaCE JAM using cubic polynomials to neighborhood selection, the  nonparanormal skeptic, and graphical random forests with stability selection. Minimizing the $\textsc{BIC}$ for SpaCE JAM yielded a graph with 16 total edges. We compared SpaCE JAM to competing methods, selecting tuning parameters  such that each resulting estimated graph contained 16 edges, as well as 10 and 20 edges for the sake of comparison. Figure~\ref{cellsfig} displays the estimated graphs, along with the directed graph presented in  \citet{sachs2005causal}.

The graphs estimated using different methods are qualitatively different. If we treat the directed graph from \citet{sachs2005causal} as the ground truth, then SpaCE JAM with 16 edges correctly identifies 12 of the edges, compared to 11, 9, and 8 using sparse partial correlation, the nonparanormal skeptic, and random forests, respectively.

Next we examined the other 13 perturbations, and found that for graphs with 16 edges, SpaCE JAM chooses on average 0.93, 0.64 and 0.2 more correct edges than sparse partial correlation, nonparanormal skeptic, and graphical random forests, respectively (p = 0.001, 0.19 and 0.68 using the paired t-test). Since graphical random forests  does not permit arbitrary specification of graph size, when graphs with 16 edges could not be obtained, we used the next largest graph.

In Section~\ref{sec:intro}, we showed that these data are not well-represented by linear models even after the nonparanormal transformation. The superior performance of SpaCE JAM in this section confirms this observation. The differences between the SpaCE JAM and graphical random forests results indicate that the approach taken for modeling non-linearity  does affect the results obtained.

\section{Extension to directed graphs}
\label{sec:dags}
In certain applications, it can be of interest to estimate the  causal relationships underlying a set of features, typically represented as a directed acyclic graph.  Though directed acyclic graph estimation is in general NP-hard, it is computationally tractable when the causal ordering is known.  In fact, in this case, a modification of  neighborhood selection is equivalent to the graphical lasso \citep{shojaie2010penalized}. We extend the penalized likelihood framework of \cite{shojaie2010penalized} to non-linear additive models  by solving
$$
\Minimize_{\beta_{jk}, 2 \leq j \leq p, k \prec j}\left \{ \frac{1}{2n} \| \x_j - \sum_{k \prec j} \bPsi_{jk} \beta_{j k }\|_2^2 + \lambda  \sum_{k \prec j}  \| \bPsi_{jk} \beta_{j k } \|_2  \right \},
$$
where $k \prec j$ indicates that $k$ precedes $j$ in the causal ordering.
When $\bPsi_{jk}=\x_k$, the model is exactly the penalized Gaussian likelihood approach of \cite{shojaie2010penalized}.

\begin{figure}[t]
\vspace{-10pt}
\includegraphics[width=\textwidth]{\detokenize{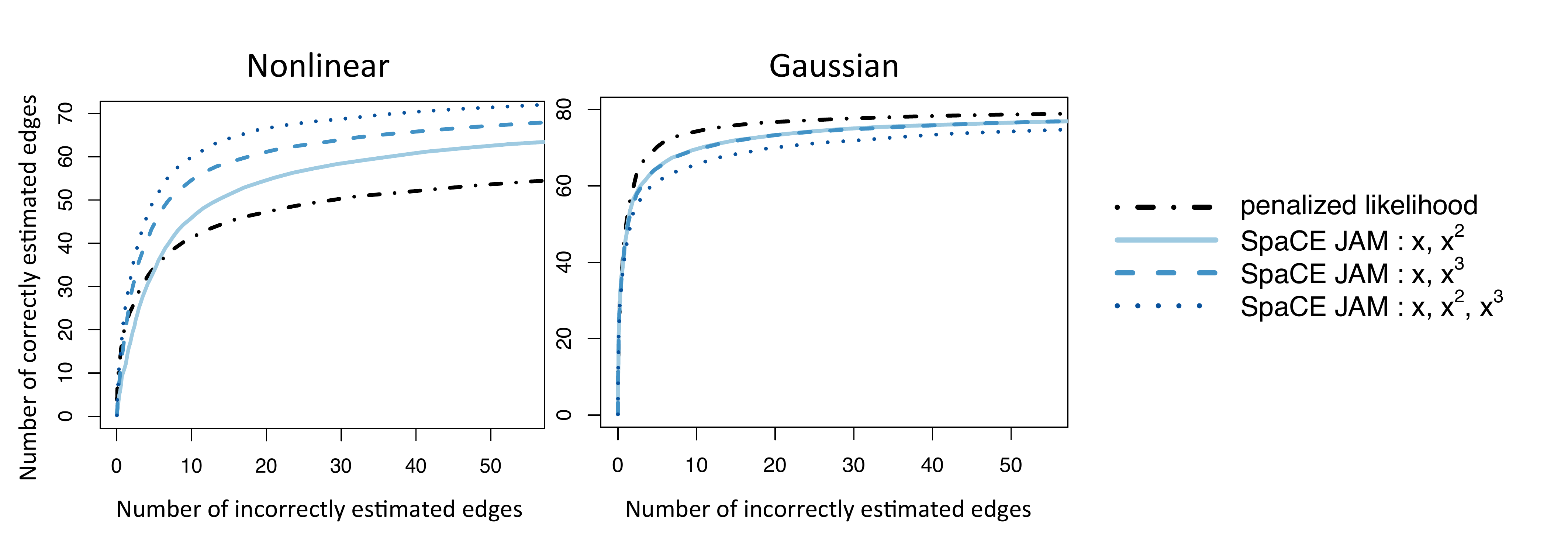}}
\caption{Simulation example with directed acyclic graphs.  The simulation is exactly as in Section~\ref{sec:sim} and Figure~\ref{simfig}. For each method, the number of correctly and incorrectly estimated edges are averaged over $100$ simulated data sets, for a range of 100 tuning parameter values. The competing method is that of \cite{shojaie2010penalized}. }
\label{dag}
\end{figure}

Figure~\ref{dag} displays the same simulation scenario as Section \ref{sec:sim}, but with the directed graph estimated using the (known) causal ordering.  Results are compared to the penalized Gaussian likelihood approach of \cite{shojaie2010penalized}. SpaCE JAM performs best when the true relationships are non-linear, and performs competitively when the relationships are linear.
\vspace{-30pt}
\section{Theoretical Results}
\label{sec:theory}
In this section, we provide theory for consistency of the SpaCE JAM graph estimate. Here, we focus on theory for undirected graphs. Similar results also hold for directed graphs, but we omit them due to space considerations. The theoretical development follows that of sparsistency results for sparse additive models with orthogonal series smoothers \citep{ravikumar2009sparse}.

First, we must define the graph for which SpaCE JAM is consistent. Recall that we have the random vector $x =(x_1, \dots, x_d) \sim \mathcal{P}$, and   ${X}  = [\, \x_{1}, \dots ,\x_{d}] \in \mathbb{R}^{n \times d}$ is a matrix where each row is an independent draw from $\mathcal{P}$.  For each $(j,k) \in V \times V$ consider the orthogonal set of basis functions $\psi_{jkt}(\cdot), \,\, t \in \mathbb{N}$. Define the population level parameters $\beta^*_{jk} \in \mathbb{R}^\infty$ as
$$\left \{ \beta^*_{jk}, k=1,\dots, d \right \} \equiv \argmin_{\beta_{jk}\, :\,  k=1, \dots, d} \left \{ \E |x_{j} - \sum_{k \neq j} \sum_{t=1}^\infty \psi_{jkt}(x_{k}) \beta_{jkt} |^2 \right\}, \quad j =1,\dots,d.$$
 Let  $S_j = \{k : \| \beta^*_{jk}\| \neq 0\}$ and $s_j=|S_j|$.
Let $f_{jk}(x_k) =  \sum_{t=1}^\infty \psi_{jkt}(x_k)\beta^*_{jkt} \in \mathcal{F}$.  Then
$$x_j = \sum_{k \in S_j}f_{jk}(x_k)+ \epsilon_j,\quad j = 1, \dots, d,$$
where $\epsilon_1,\dots, \epsilon_d$ are  residuals, and $\sum_{k \in S_j}f_{jk}(x_k)$ is the best additive approximation to $\E[x_j \mid \{x_k : k \neq j\}]$, in the least-squares sense. We wish to determine which of the $f_{jk}(\cdot)$ are zero.

On observed data, we use a finite set of basis functions to model the $f_{jk}(\cdot)$.
Denote the set of $r$ orthogonal basis functions used in the regression of $\x_j$ on $\x_k$ as $\bPsi_{jk} = [\psi_{jk1}(\x_k), \dots, \psi_{jkr}(\x_k)]$, a matrix of dimension $n \times r$ such that $\bPsi_{jk}^T\bPsi_{jk}/n =I_ r$.  Let $\beta^{*(r)}_{jk} = [\beta^*_{jk1}, \dots, \beta^*_{jkr}]^T$ denote the first $r$ components of $\beta^*_{jk}$. Further, let $\bPsi_{S_j}$ be the concatenated basis functions in $\{\bPsi_{jk} : k \in S_j \}$, thus $\bPsi_{S_j}$ is a matrix of dimension $n \times s_j r$. Also let $\bSigma_{S_j,S_j} = \left(\frac{1}{n}\bPsi_{S_j}^T \bPsi_{S_j}\right)$ and $\bSigma_{jk, S_j} =  \left(\frac{1}{n}\bPsi_{jk}^T \bPsi_{S_j}\right)$.
Define the sub-gradient of the penalty  in $(\ref{eqn:us})$ with respect to $ \beta_{jk}$ as $g_{jk}(\beta)$.  On the set $S_j$, we write the concatenated sub-gradients as $g_{S_j}$,
 a vector of length $s_jr$.

Let $\hat \beta$ be the parameter estimates from solving \eqref{eqn:us},  let $\hat E_n = \{ (j,k) : \|\hat \beta_{jk}\|_2^2 + \|\hat \beta_{kj}\|_2^2 \neq 0\}$ be the corresponding estimated edge set, and let $E^* = \{ (j,k) : \, k \in S_j \text{ or } j \in S_k \}$ be the graph obtained from the population level parameters.  In Theorem~\ref{thm1}, we give precise conditions under which $\mathrm{pr}(\hat E_n = E^*) \rightarrow 1$ as $n \rightarrow \infty$.
\vspace{-15pt}
\begin{theorem}\label{thm1}
Let the functions $f_{jk}$ be sufficiently smooth, in the sense that if $f^{(r)}_{jk} = \sum_{t=1}^r \psi_{jkt}(x_k)\beta^*_{jkt}$, then $|f_{jk}^{(r)}(x_{k}) - f_{jk}(x_{k})| = O_p(1/r^m)$
uniformly in $(j,k) \in V \times V$ for some $m \in \mathbb{N}$.  For $j =1 , \dots, d,$ assume the basis functions satisfy
$\Lambda_{min}( \bSigma_{S_j,S_j} ) \geq C_{min} > 0 $
with probability tending to 1.
Assume the irrepresentability condition,
\begin{equation} \label{irrep1} \|\bSigma_{jk, S_j} \bSigma_{S_j,S_j} ^{-1}\hat g_{S_j}\|_2^2 +\|\bSigma_{kj, S_k} \bSigma_{S_k,S_k} ^{-1}\hat g_{S_k}\|_2^2 \leq 1 - \delta, \end{equation}
holds for $(j,k) \notin E^*$ and some $\delta > 0$ with probability tending to 1, where $\hat g_{S_j} = g_{S_j}(\hat \beta)$. Assume the following conditions on the number of edges $|E^*|$, the neighborhood size $s_j$, the regularization parameter $\lambda$, and the truncation dimension $r$:
$$\frac{r \log(r|E^{*c}|)}{\lambda^2n} \rightarrow 0,  \quad \max_j\frac{r s_j \log(r|E^*|) }{\lambda^2 n} \rightarrow 0, \quad  \max_j \frac{s_j}{r^m \lambda}\rightarrow 0 ,\quad \mathrm{and}$$
$$ \frac{1}{ \rho^*} \max_j \left [  \left(\frac{s_j r\log(r|E^*|)}{n}\right)^{1/2}+ \frac{s_j}{r^{m}} +\lambda(rs_j)^{1/2} \right] \rightarrow 0$$
where $\rho^* = \min_j \min_{k \in S_j} \|\beta^*_{jk}\|_\infty$.  Further, assume the variables
\begin{align*}
\xi_{jkt} & \equiv  \psi_{jkt}(x_{k})\epsilon_{j} \quad \mathrm{for}\, j,k \in V,\, \mathrm{and}\, j =1, \dots, d
 \end{align*}
have exponential tails, that is $\mathrm{pr}[\,|\xi_{jkt}| > z ] \leq a e^{-bz^2}$ for some $a, b > 0$.

Then, the SpaCE JAM graph estimate is consistent: $pr(\hat E_n = E^*) \rightarrow 1$ as $n \rightarrow \infty$.
\vspace{-20pt}
\end{theorem}

\section{Extension of SpaCE JAM to high dimensions}
\label{sec:extension}
In this section, we propose {an approximation to SpaCE JAM that can speed up computations in high dimensions}. Our proposal is motivated by recent work in the Gaussian setting by \citet{witten2011} and \citet{mazumder2012exact}.  They showed that for the graphical lasso (\ref{glasso}), the connected components of the estimated conditional independence graph  are precisely the connected components of the estimated marginal independence graph, where the $j$th and $k$th variables are considered marginally independent when $|\x_j^T\x_k| < \lambda$.  Consequently, one can obtain the exact solution to the graphical lasso problem in substantially reduced computational time by identifying the connected components of the marginal independence graph, and solving the graphical lasso optimization problem on the variables within each connected component.

We now apply the same principle to SpaCE JAM in order to quickly approximate the solution to (\ref{eqn:us}) in high dimensions. Let $\rho_m^{(jk)} = \sup_{f,g \in \mathcal{F}} \rho(f(x_k), g(x_j))$ be the maximal correlation between $x_j$ and $x_k$ over the univariate functions in $\mathcal{F}$ such that $f(x_k)$ and $g(x_j)$ have finite variance.  Define the marginal dependence graph $\Gamma_M = (V, E_M)$, where $(j,k) \in E_M$ when $\rho_m^{(jk)} \neq 0$.  If the $j$th and $k$th variables are in different connected components of $\Gamma_M$, then they must be conditionally independent in the large-sample SpaCE JAM graph. Theorem~\ref{thm:screen}, proven in the Appendix, makes this assertion precise.
\begin{theorem}
\label{thm:screen}
Let $C_1, \dots C_l$ be the connected components of $\Gamma_M$.  Suppose  the space of functions $\mathcal{F}$ contains linear functions. If $j \in C_u$ and $k \notin C_u$ for some $1 \leq u \leq l$, then $(j,k) \notin E^*$.
\end{theorem}
 Theorem~\ref{thm:screen} forms the basis for Algorithm 2. %, which allows us to screen edges based on pairwise association in order to achieve computational gains.
  There, we approximate the maximal correlation using the canonical correlation \citep{mardia1980multivariate} between the basis expansions $\bPsi_{kj}$ and $\bPsi_{jk}$: $\hat \rho_{m}^{(jk)} = \max_{v,w \in \mathbb{R}^r} \rho(\bPsi_{jk}v, \bPsi_{jk}w)$.

% but rather than using Pearson correlation to quantify marginal dependence between a pair of variables, we use the canonical correlation between their basis expansions \citep{mardia1980multivariate}.  The resulting procedure is given in Algorithm~\ref{alg:screen}.  We show that in large samples, this procedure results in computational gains without breaking up any truly connected components of the conditional independence graph. More specifically, d

\begin{algorithm}[h]
\caption{A fast approximation for SpaCE JAM in high dimensions}\label{alg:screen}
\vspace*{-12pt}
\begin{tabbing}
\enspace 1: For $(j,k) \in V \times V$, calculate $\hat \rho_m^{(jk)}$, the sample canonical correlation between $\bPsi_{kj}$ and $\bPsi_{jk}$. \\
\enspace 2: Construct the marginal independence graph:  $(j,k) \in \hat \Gamma_M$ when $|\hat \rho_m^{(jk)}| \geq \lambda_2$. \\
\enspace 3: Find the connected components $C_1, \dots C_l$ of $\hat \Gamma_M$.\\
\enspace 4: Perform Algorithm 1 on each connected component.
\end{tabbing}
\vspace{-10pt}
\end{algorithm}
%:
In order to show that i) Algorithm~\ref{alg:screen} provides an accurate approximation to the original SpaCE JAM problem, ii) the resulting estimator outperforms methods that rely on Gaussian assumptions when those assumptions are violated, and iii) Algorithm~\ref{alg:screen} is indeed faster than Algorithm 1, we replicated the graph  used in Section \ref{sec:sim} five times. This gives  $d = 500$ variables, broken into five components. We took $n=250$, and set $\bPsi_{jk} = [\, \x_k, \x_k^2, \x_k^3\,]$.

\begin{figure}[h]
\centering
\vspace{-20pt}
\includegraphics[width=0.5\textwidth]{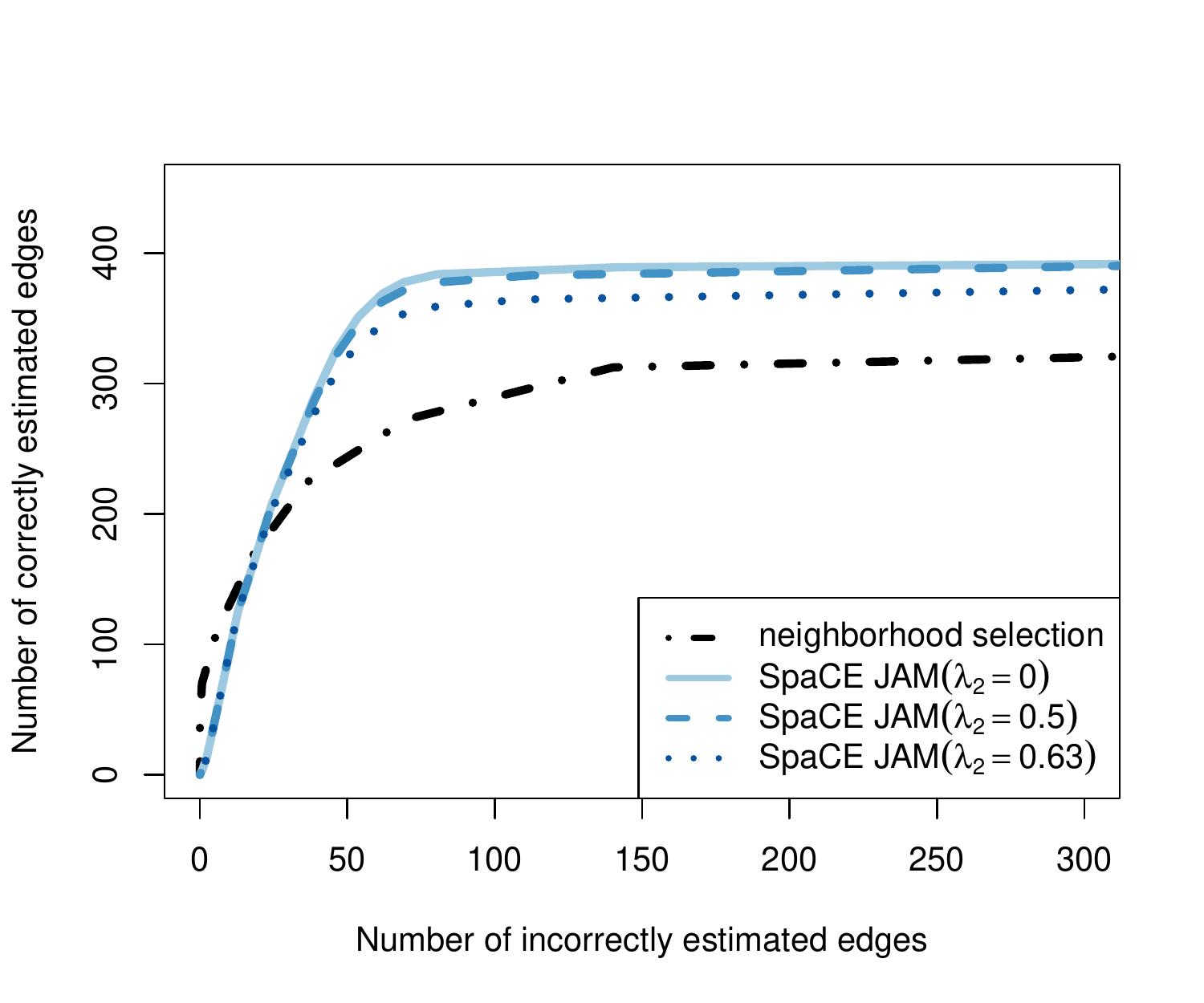}
\caption{Performance of SpaCE JAM using Algorithm~\ref{alg:screen}. The number of correctly and incorrectly estimated edges are averaged over 100 simulated data sets, for each of 100 tuning parameter values.  SpaCE JAM was applied using cubic polynomials as basis functions. The competing method is that of \cite{meinshausen2006}.}
\label{fig:screen}
\vspace{-10pt}
\end{figure}

In Figure~\ref{fig:screen} we see that when $\lambda_2$ in Algorithm~\ref{alg:screen} is small, there is little loss in statistical efficiency relative to the full SpaCE JAM algorithm (Algorithm 1), which is a special case of Algorithm~\ref{alg:screen} with $\lambda_2 = 0$.  Further, we see that  SpaCE JAM outperforms neighborhood selection even when $\lambda_2$ is large. Using Algorithm 2 with $\lambda_2 = 0.5$ and $\lambda_2=0.63$ led to a reduction in computation time over Algorithm 1 by $25\%$ and $70\%$, respectively.

We note here that Theorem \ref{thm:screen} continues to hold if maximal correlation $\rho_m^{(jk)}$ is replaced with some other measure of  marginal association $\rho_*^{(jk)}$, provided that $\rho_*^{(jk)}$ dominates maximal correlation in the sense that $\rho_*^{(jk)} = 0$ implies that $\rho_m^{(jk)} = 0$. {That is, any measure of marginal association, such as mutual information, which detects the same associations as maximal correlation (i.e. $\rho_*^{(jk)} \neq 0$ if  $\rho_m^{(jk)} \neq 0$) can be used in Algorithm 2.}

\vspace{-20pt}
\section{Discussion}
\label{sec:discussion}
In this paper we have discussed conditional independence graph estimation for non-normal data.  In the high-dimensional setting, assumptions on the joint distribution of a  set of variables cannot reasonably be expected to hold, and cannot be checked.
  Therefore, we have proposed SpaCE JAM, which models conditional distributions using flexible additive models, and thereby gives more accurate graph estimation for non-normal data. {The} \verb=R= {package} \verb!spacejam! at \verb!cran.r-project.org/package=spacejam! {implements the proposed approach}.

{A possible extension to this work involves  accommodating temporal information. We could take advantage of the natural ordering induced by time, as considered by \cite{shojaie2010discovering}, and apply SpaCE JAM for directed graphs. We leave this to future work.}

\section*{Acknowledgments}
We thank two anonymous reviewers and an associate editor for helpful comments, Thomas Lumley for valuable insights, and Han Liu and Bernd Fellinghauer for providing \verb=R= code.

\appendix

\appendixone
\setstretch{1}
\section*{Appendix 1: Technical proofs}

\subsection{Proof of Theorem 1}
First, we restate a theorem which will be useful in the proof of the main result.
\begin{theorem}(Kuelbs \& Vidyashankar 2010)
Let $\{ \xi_{n,j,i}:\, i = 1, \dots, n, j \in A_n \}$ be a set of random variables such that $\xi_{n,j,i}$ is independent of $\xi_{n,j,i'}$ for $i \neq i'$.  That is, $\xi_{n,j,i}, i = 1, \dots, n$ denotes independent observations of feature $j$, and the features are indexed by some finite $A_n$. Assume $\E [\xi_{n,j,i}] = 0$, and there exist constants $a > 1 $ and $b > 0 $ such that $\mathrm{pr}(|\xi_{n,j,i}| \geq x) \leq a e^{-b x^2}$ for all $x > 0$.  Further, assume that $|A_n| < \infty$ for all $n$ and that $|A_n| \rightarrow \infty$ as $n \rightarrow \infty$.  Denote $z_{n,j} = \sum_{i=1}^n \xi_{n,j,i}$.  Then
$$\frac{\max_{j \in A_n}|z_{n,j}|}{n} = O_p\left(\left(\frac{\log(|A_n|)}{n}\right)^{1/2} \right).$$
\end{theorem}
We now prove Theorem 1.
\begin{proof}
First, $\hat \beta$ is a solution to (\ref{eqn:us}) if and only if
\begin{equation}\label{subgrad} -\frac{1}{n}\bPsi_{jk}^T\left(\x_j - \sum_{l \neq j}\bPsi_{jl} \hat \beta_{jl}\right) + \lambda g_{jk}(\hat \beta) = 0 \quad  \text{ for } (j,k) \in V \times V,\end{equation}
where $g_{jk}(\hat \beta)$ is the vector satisfying

\begin{align*}
g_{jk}(\beta) =  \frac{\bPsi_{jk}\beta_{jk}}{(\|\bPsi_{jk}\beta_{jk}\|_2^2 + \|\bPsi_{kj}\beta_{kj}\|_2^2)^{1/2}} \quad &\text{ when } \quad \|\beta_{jk}\|_2 + \|\beta_{kj}\|_2 \neq 0 \\
\|g_{jk}(\beta)\|_2^2 + \|g_{kj}(\beta)\|_2^2 \leq 1 \quad &\text{ when }\quad  \|\beta_{jk}\|_2 + \|\beta_{kj}\|_2 = 0.
\end{align*}
We base our proof on the primal-dual witness method of \citet{wainwright2009sharp}.  That is, we construct a coefficient-subgradient pair $(\hat \beta, \hat g)$, and show that they solve \eqref{eqn:us} and produce the correct sparsity pattern, with probability tending to 1.
For $(j,k) \in E^*$, we construct $\hat \beta_{jk}$ and the corresponding sub-gradients $\hat g_{jk}$ using SpaCE JAM, restricted to edges in $E^*$:
\begin{equation}
\argmin_{\beta_{jk}: (j,k) \in E^*} \left \{ \frac{1}{2n} \sum_{j=1}^d \| \x_j - \sum_{k \in S_j} \bPsi_{jk} \beta_{jk} \|_2^2 + \lambda \sum_{(j,k) \in E^*} \left( \| \bPsi_{jk} \beta_{jk} \|_2^2 + \|\bPsi_{kj} \beta_{jk}\|_2^2  \right)^{1/2} \right \}.
\end{equation}
For $(j,k) \in E^{*c}$, we set $\hat \beta_{jk} = 0$, and use \eqref{subgrad} to solve for the remaining $\hat g_{jk}$ when $k \notin S_j$.  Now, $\hat \beta$ is a solution to \eqref{eqn:us} if
\begin{align}
\label{cond2} &\|g_{jk}(  \hat \beta_{jk})\|_2^2 + \| g_{kj}(  \hat \beta_{kj})\|_2 \leq 1 \text{ for } (j,k) \notin E^*.
\end{align}
In addition, $\hat E_n = E^*$ when
\begin{align}
\label{cond1} &\hat \beta_{S_j} \neq 0 \text{ for } j = 1, \dots , d.
\end{align}
Thus, it suffices to show that that Equations~\eqref{cond2} and \eqref{cond1} hold with high probability.
\newline

\noindent \underline{Condition (\ref{cond1})}: We start with the `primal' problem.  The stationary condition for $\hat \beta_{S_j}$ is given by
$$-\frac{1}{n} \bPsi_{S_j}^T(\x_j - \bPsi_{S_j}\hat \beta_{S_j}) + \lambda \hat g_{S_j} = 0.$$
Denote by $\sum_{k \in S_j}\left[ f_{jk}(\x_j) -f^{(r)}_{jk}(\x_j)  \right] = { w}_j$ the truncation error from including only $r$ basis terms.  We can write $\x_j = \bPsi_{S_j}\beta^{*(r)}_{S_j} + {w}_j +{ \epsilon}_j$.  And so
$$\frac{1}{n} \bPsi_{S_j}^T\left (\bPsi_{S_j} (\hat \beta_{S_j} - \beta^{*(r)}_{S_j}) - {w}_j -{ \epsilon}_j\right) + \lambda \hat g_{S_j} = 0,$$
or
\begin{equation} \label{eqn:beta}
(\hat \beta_{S_j} - \beta^{*(r)}_{S_j}) = \left(\frac{1}{n}\bPsi_{S_j}^T\bPsi_{S_j} \right)^{-1}\left (\frac{1}{n}\bPsi_{S_j}^T { w}_j + \frac{1}{n}\bPsi_{S_j}^T { \epsilon}_j - \lambda \hat g_{S_j}\right),
\end{equation}
using the assumption that $\frac{1}{n}\bPsi_{S_j}^T\bPsi_{S_j}$ is invertible. We will now show that the inequality
\begin{equation}\label{eqn:minbeta}
\max_j \|\hat \beta_{S_j} - \beta^{*(r)}_{S_j}\|_\infty < \min_j \min_{k \in S_j} \|\beta^{*(r)}_{jk}\|_\infty/2 \equiv \rho^*/2
\end{equation}
holds with high probability.  This implies that $\|\hat \beta_{jk}\|_2 \neq 0$ if $\|\beta_{jk}^{*(r)}\|_2 \neq 0$.

From \eqref{eqn:beta} we have that
\begin{align*}
\max_j \|\hat \beta_{S_j} - \beta^{*(r)}_{S_j}\|_\infty &\leq \max_j  \left \| \bSigma_{S_j,S_j} ^{-1} \frac{1}{n}\bPsi_{S_j}^T { w}_j \right \|_\infty +\max_j  \left\| \bSigma_{S_j,S_j} ^{-1} \frac{1}{n}\bPsi_{S_j}^T{\epsilon}_j \right\|_\infty + \max_j \lambda \left\| \bSigma_{S_j,S_j} ^{-1} \hat g_{S_j} \right\|_\infty \\
&\equiv T_1 + T_2 + T_3.
\end{align*}
Thus, to show \eqref{eqn:minbeta} it suffices to bound $T_1$, $T_2$, and $T_3$.
\vspace{-25pt}
\begin{itemize}
\item Bounding $T_1$:

By assumption, we have that $| f_{jk}^{(r)}(x_{k}) - f_{jk}(x_{k})| = O_p(1/r^m)$ uniformly in $k$.  Thus, $n^{-1/2}\|{w}_j\|_2 = \left \|1/n \sum_{k \in S_j}\left[ f_{jk}^{(r)}(\x_{k}) - f_{jk}(\x_{k})\right] \right \|_2 = O_p(s_j/r^m)$ uniformly in $j$.

This implies that
\begin{align*}
T_1 &\leq \max_j  \left \|  \bSigma_{S_j,S_j} ^{-1} \frac{1}{n}\bPsi_{S_j}^T {w}_j \right \|_2  \leq \max_j \left \| \bSigma_{S_j,S_j} ^{-1}\frac{1}{\surd{n}}\bPsi^T_{S_j}\right \|_2  \frac{1}{\surd{n}} \left\| { w}_j \right \|_2\\
& \leq C^{-1/2}_{min}   \max_j O_p\left(s_j/r^m\right) = O_p\left(\frac{\max_j s_j }{r^{m}}\right).
\end{align*}
In the above, we used that $\Lambda_{max}\left(\bSigma_{S_j,S_j} ^{-1}\frac{1}{\surd{n}}\bPsi^T_{S_j}\right) = \left(\Lambda_{min}\left(\bSigma_{S_j,S_j}\right)\right)^{1/2}$.
\item Bounding $T_2$:

Here, we use Theorem A1 which bounds the $\ell_\infty$ norm of the average of high-dimensional i.i.d. vectors.
 First, by the definition of $\epsilon_j$ we must have that $\E [\,\psi_{jkt}(x_k)\epsilon_j\,]  = 0$, i.e. the residuals are uncorrelated with the covariates.

Let $z_{jkt} \equiv \psi_{jkt}(\x_k)^T{\epsilon}_j$, which is sum of $n$ independent random variables with exponential tails. We have that
$$\max_j \|\bPsi_{S_j}^T { \epsilon}_j\|_{\infty}/n = \max_j\max_{k \in S_j} \max_{t=1,\dots,r} |z_{jkt}|/n \leq \max_{(j,k) \in  E^*}\max_{t=1,\dots,r}\left\{|z_{jkt}|\vee|z_{kjt}|/n\right\},$$
the maximum of $2r|E^*|$ elements.  We can thus apply Theorem A1, with $A_n$ indexing the $2r|E^*|$ elements above, to obtain
\begin{align*}
T_2  &=  \max_j  \left\| \bSigma_{S_j,S_j} ^{-1} \frac{1}{n}\bPsi_{S_j}^T{\epsilon}_j \right\|_\infty \leq  \max_j  \left\| \bSigma_{S_j,S_j} ^{-1}\right \|_\infty \left\| \frac{1}{n}\bPsi_{S_j}^T{\epsilon}_j \right\|_\infty \\
& \leq \max_j (rs_j)^{1/2} C^{-1}_{min} O_p\left(\left(\frac{ \log(2r|E^*|)}{n}\right)^{1/2}\right)= O_p\left(\left(\frac{\max_js_j r\log(r|E^*|)}{n}\right)^{1/2}\right).
\end{align*}

\item Bounding $T_3$:

We have that $\|\hat g_{jk}\|_2^2 \leq 1$ for $(j,k) \in E^*$, so
\begin{align*}
T_3 &\leq \lambda \max_j\left \|  \bSigma_{S_j,S_j} ^{-1} \right\|_{\infty} \leq \lambda \max_j (rs_j)^{1/2} \left \| \bSigma_{S_j,S_j} ^{-1} \right\|_{2} \leq \lambda \max_j\frac{(rs_j)^{1/2}}{C_{min}}.
\end{align*}

Altogether, we have shown that
$$ \max_j \|\hat \beta_{S_j} - \beta^{*(r)}_{S_j}\|_\infty \leq  O_p\left(\frac{\max_j s_j}{r^{m}}\right) +   O_p\left(\left(\frac{(\max_js_j )r\log(r|E^*|)}{n}\right)^{1/2}\right) + \lambda \max_j\frac{(rs_j)^{1/2}}{C_{min}}.$$
By assumption,
$$ \frac{1}{ \rho^*} \max_j \left [  \left(\frac{s_j r\log(r|E^*|)}{n}\right)^{1/2}+ \frac{s_j}{r^{m}} +\lambda(rs_j)^{1/2} \right] \rightarrow 0$$
which implies that $\max_j \|\hat \beta_{S_j} - \beta^{*(r)}_{S_j}\|_\infty < \rho^*/2$ with probability tending to 1 as $n \rightarrow \infty$.
\end{itemize}

\vspace{20pt}
\noindent \underline{Condition (\ref{cond2})}:
We now consider the `dual' problem.  That is, we must show that $\| \hat g_{jk}\|_2 +  \| \hat g_{kj}\|_2 \leq 1$ for each $(j,k) \notin E^*$.  From the discussion of Condition (\ref{cond1}), we know that
\begin{align*}
 \hat g_{jk} &=\frac{1}{\lambda n} \bPsi_{jk}^T\left (\bPsi_{S_j} (\hat \beta_{S_j} - \beta^{*(r)}_{S_j}) - {w}_j -{ \epsilon}_j\right) \\
 & =\frac{1}{\lambda n} \bPsi_{jk}^T\left (\bPsi_{S_j} \bSigma_{S_j,S_j}^{-1}\left (\frac{1}{n}\bPsi_{S_j}^T  { w}_j + \frac{1}{n}\bPsi_{S_j}^T{ \epsilon}_j - \lambda \hat g_{S_j}\right) -  {w}_j- { \epsilon}_j\right) \\
 &= -\frac{1}{\lambda n} \bPsi_{jk}^T\left ({  I} - \frac{1}{n}\bPsi_{S_j} \bSigma_{S_j,S_j}^{-1}\bPsi_{S_j}^T\right){ w}_j - \frac{1}{\lambda n} \bPsi_{jk}^T\left ({ I} -\frac{1}{n} \bPsi_{S_j}  \bSigma_{S_j,S_j}^{-1}\bPsi_{S_j}^T\right){ \epsilon}_j \\
 & \qquad - \frac{1}{ n} \bPsi_{jk}^T\bPsi_{S_j} \bSigma_{S_j,S_j}^{-1}\hat g_{S_j} \\
& \equiv M_1^{jk}+ M_2^{jk} + M_3^{jk}.
\end{align*}
We will proceed by bounding $\|M^{jk}_1\|_2 + \|M^{kj}_1\|_2$, $\|M^{jk}_2\|_2 + \|M^{kj}_2\|_2$ and $\|M^{jk}_3\|_2 + \|M^{kj}_3\|_2$, which will give us a bound for the quantity of interest, $\| \hat g_{jk}\|_2 +  \| \hat g_{kj}\|_2$.

\begin{itemize}

\item Bounding $M_1$:

When bounding $T_1$ earlier, we saw that $n^{-1/2}\|{ w}_j\|_2 = O_p(s_j/r^m)$.  Now $\left (I - \bPsi_{S_j} \bSigma_{S_j,S_j}^{-1}\bPsi_{S_j}^T/n\right)$ is a projection matrix, and by design $n^{-1/2}\bPsi_{jk}$ is orthogonal, so that all the eigenvalues of $n^{-1/2}\bPsi_{jk}$ are 1. Therefore

\begin{align*}
\| M_1^{jk}\|_2& \leq \frac{1}{\lambda }n^{-1/2}\|\bPsi_{jk}\|_2 \, n^{-1/2}\left \|  { w}_j \right \|_{2} = O_p\left(\frac{s_j}{\lambda r^m}\right),
\end{align*}
and
$$\| M_1^{jk}\|_2 + \| M_1^{kj}\|_2 \leq O_p\left(\frac{s_j \vee s_k }{\lambda r^m}\right),$$
which tends to zero because
$\frac{s_j }{\lambda r^m} \rightarrow 0$
uniformly in $j$.
\item Bounding $M_2$:

First, note that

\begin{align*}
 \lambda \| M^{jk}_2\|_2 &\leq n^{-1}\|\bPsi_{jk}^T{\epsilon}_j\|_2 +  n^{-1/2}\|\bPsi_{jk} \|_2 \left\|n^{-1/2}\bPsi_{S_j} \bSigma_{S_j,S_j}^{-1}\right\|_2\|\bPsi^T_{S_j} { \epsilon}_j\|_2/n \\
 &\leq  n^{-1}\|\bPsi_{jk}^T\epsilon_j\|_2 + C^{-1/2}_{min}\|\bPsi^T_{S_j} { \epsilon}_j\|_2/n\\
& \leq  \surd{r} \|\bPsi_{jk}^T\epsilon_j\|_\infty/n + \left(rs_j /C_{min}\right)^{1/2}\|\bPsi^T_{S_j} {\epsilon}_j\|_\infty/n.
\end{align*}
Then, applying Theorem A1, as in the bound for $T_2$, we get
\begin{align*}
\lambda\max_{(j,k) \in  E^{*c}} \| M^{jk}_2\|_2 \leq O_p\left(\left(\frac{r \log(r| E^{*c}|) }{n}\right)^{1/2}\right) + O_p\left(\left(\frac{r \max_j s_j \log(r|E^{*}|) }{n}\right)^{1/2}\right).
\end{align*}
Thus, $\max_{(j,k) \in E^{*c}} \left\{ \| M^{jk}_2\|_2+\| M^{jk}_2\|_2\right\} \rightarrow 0$ when
$$\frac{r \log(r| E^{*c}|) }{\lambda^2 n} \rightarrow 0 \quad \text{ and } \quad \max_j\frac{r s_j \log(r|E^*|) }{\lambda^2 n} \rightarrow 0.$$

\item Bounding $M_3$:

By the irrepresentability assumption, we have that $\| M_3^{jk}\|^2_2 + \| M_3^{kj}\|^2_2 \leq 1 - \delta$ with probability tending to 1.
\end{itemize}

Thus, since $\|M^{jk}_1\|_2 + \|M^{kj}_1\|_2 + \|M^{jk}_2\|_2 + \|M^{kj}_2\|_2 = o_p(1)$, we have that for each $(j,k) \in E^{*c}$
$$\max_{(j,k) \in E^{*c}}\left\{ \| \hat g_{jk}\|_2 +  \| \hat g_{kj}\|_2 \right\}  \leq 1-\delta$$
with probability tending to 1. Further, since we have strict dual feasibility, i.e. $\| \hat g_{jk}\|_2 +  \| \hat g_{kj}\|_2 < 1 $  for $(j,k) \in E^{*c}$, with probability tending to 1, the estimated graph is unique.
\end{proof}

\subsection{Proof of Theorem 2}
\begin{proof}
Consider a variable $j$, with $j \in C_u$.  Our large-sample model requires minimizing $\E|x_{j} - \sum_{k \neq j}  \sum_{t=1}^\infty \psi_{jkt}(x_{k}) \beta_{jkt} |^2$ with respect to the $\beta_{jkt}$, or equivalently, minimizing $$\E|x_j - \sum_{k \neq j} f_{jk}(x_k)|^2 $$ over functions $f_{jk} \in \mathcal{F}$. We have that

\begin{align*}
\E|x_j -& \sum_{k \neq j} f_{jk}(x_k)|^2 = \E x_j^2 -2\sum_{k\neq j} \E[\, x_j f_{jk}(x_{k}) \,]+ \sum_{k\neq j}\sum_{l \neq j} \E[\,f_{jk}(x_k)f_{jl}(x_l)\,]  \\
  & =  \E x_j^2 -2\sum_{k \in C_u} \E[\, x_j f_{jk}(x_{k}) \,] -2\sum_{k \notin C_u} \E[\, x_j f_{jk}(x_{k}) \,] +  \sum_{k \in C_u}\sum_{ l \in C_u}  \E[\,f_{jk}(x_k)f_{jl}(x_l)\,] \\&+\sum_{k\notin C_u}\sum_{l \notin C_u}  \E[\,f_{jk}(x_k)f_{jl}(x_l)\,] + 2\sum_{k\notin C_u}\sum_{l \in C_u}  \E[\,f_{jk}(x_k)f_{jl}(x_l)\,] .\\
\end{align*}
By assumption $\sum_{k \notin C_u} \E[\, x_j f_{jk}(x_{k}) \,] = \sum_{k\notin C_u} \sum_{l \in C_u}  \E[\,f_{jk}(x_k)f_{jl}(x_l)\,]=0$.  Thus, collecting terms, we get
$$ \E|x_j - \sum_{k \neq j} f_{jk}(x_k)|^2= \E |x_j - \sum_{k\in C_u} f_{jk}(x_k)|^2 + \E |\sum_{k\notin C_u}f_{jk}(x_k)|^2.$$
Minimization of this quantity with respect to $\{ f_{jk} \in \mathcal{F}, k \notin C_u \}$ only involves the last term, which achieves its minimum at zero when $ f_{jk}(\cdot) = 0$ almost everywhere for each $k \notin C_u$.

\end{proof}

\bibliographystyle{biometrika}
%\bibliography{/Users/arie/Dropbox/NonLinearGraphicalLasso/Ariesrefs}
\bibliography{spacejam_biom_arxiv.bbl}

\end{document}